\documentclass[twocolumn,showpacs,preprintnumbers,floatfix,superscriptaddress,amsmath,amssymb]{revtex4-1}

\usepackage{graphicx}
\usepackage{dcolumn}
\usepackage{bm}
\usepackage{hyperref}
\usepackage[rightcaption]{sidecap}
\usepackage{float}
\hypersetup{
  colorlinks   = true,	
  urlcolor     = blue,	
  linkcolor    = blue,	
  citecolor   = blue	
}
\newcommand{\ohm}{$\Omega$}
\begin{document}


\title{Transition from homogeneous to inhomogeneous limit cycles: Effect of local filtering in coupled oscillators} 


\author{Tanmoy Banerjee}%
\email{tbanerjee@phys.buruniv.ac.in}
\affiliation{Chaos and Complex Systems Research Laboratory, Department of Physics, University of Burdwan, Burdwan 713 104, West Bengal, India}
\author{Debabrata Biswas} \thanks{Equal contribution}
\affiliation{Department of Physics, Rampurhat College, Birbhum 731224, West Bengal, India}
\author{Debarati Ghosh} \thanks{Equal contribution}
\affiliation{Chaos and Complex Systems Research Laboratory, Department of Physics, University of Burdwan, Burdwan 713 104, West Bengal, India}
\author{Biswabibek Bandyopadhyay} 
\affiliation{Chaos and Complex Systems Research Laboratory, Department of Physics, University of Burdwan, Burdwan 713 104, West Bengal, India}
\author{J\"urgen Kurths}
\affiliation{Potsdam Institute for Climate Impact Research, Telegraphenberg, D-14415 Potsdam, Germany}
\affiliation{Institute of Physics, Humboldt University Berlin, D-12489 Berlin, Germany}
\affiliation{Institute of Applied Physics of the Russian Academy of Sciences, 603950 Nizhny Novgorod, Russia}%


\received{:to be included by reviewer}
\date{\today}

\begin{abstract}
We report an interesting symmetry-breaking transition in coupled identical oscillators, namely the continuous transition from homogeneous to inhomogeneous limit cycle oscillations. The observed transition is the oscillatory analog of the Turing-type symmetry-breaking transition from amplitude death (i.e., stable homogeneous steady state) to oscillation death (i.e., stable inhomogeneous steady state). This novel transition occurs in the parametric zone of occurrence of rhythmogenesis and oscillation death as a consequence of the presence of local filtering in the coupling path. We consider paradigmatic oscillators, such as Stuart-Landau and van der Pol oscillators under mean-field  coupling with low-pass or all-pass filtered self-feedback and through a rigorous bifurcation analysis we explore the genesis of this transition. Further, we experimentally demonstrate the observed transition, which establishes its robustness in the presence of parameter fluctuations and noise.
\end{abstract}

\pacs{05.45.--a, 05.45.Xt}
\keywords{Amplitude death, oscillation death, Turing bifurcation, mean-field coupling}

\maketitle 

\section{Introduction}
\label{sec:intro}
Cooperative phenomena in coupled oscillators have been an active topic of extensive research in the field of physics, biology, engineering, and social science \cite{sync}. Coupled oscillators show several cooperative behaviors such as, synchronization, phase-locking, and oscillation quenching \cite{piko}. In this context the spontaneous symmetry-breaking transition from a stable homogeneous steady state (HSS) [also known as the amplitude death (AD) state] to a stable inhomogeneous steady state (IHSS) [also known as the oscillation death (OD) state] discovered by \citet{kosprl} has been in the center of recent interest. They considered two diffusively coupled Stuart-Landau oscillators with parameter mismatch and established that the symmetry-breaking transition from AD to OD state 
is equivalent to the Turing type bifurcation \cite{turing} occurs in spatially extended systems. Later, this transition has also been observed under several coupling schemes in coupled {\it identical} oscillators \cite{kospre,kurthpre,scholl4,dana,*dana1,tanpre1,*tanCD,tanpre3,mmd,srimali-sr} (see \cite{kosprep} for an elaborate review). It was also experimentally observed in coupled electronic oscillators \cite{tanpre2}.

The above mentioned AD-OD transition is due to the symmetry breaking in {\em steady states}. Therefore, the next natural question arises if there exists a similar symmetry-breaking transition in limit cycle (LC) oscillations, also? More specifically, we are interested to explore the transition from  a {\it stable} homogeneous limit cycle (HLC) to a {\it stable} inhomogeneous limit cycle (IHLC)  (or vice versa) in coupled {\em identical} oscillators. Identification and understanding of this transition is important as it may shed light on the genesis of another significant symmetry-breaking state, namely the amplitude chimera \cite{scholl_CD,*lr16}, which is the spatiotemporal coexistence of (unstable) IHLC and HLC in a network of coupled {\it identical} oscillators. Further, it may improve our understanding of various biological processes, like cellular differentiations \cite{cell2,*stem,*cell}, where a transition occurs from homogeneity to inhomogeneity. Earlier, \citet{kosprl} observed a transition from HLC to IHLC in Stuart-Landau oscillators under diffusive coupling but that transition essentially resulted from the parameter mismatch; also, it occurs around the homogeneous steady state and has no connection with the symmetry-breaking branches of OD. To the best of our knowledge the symmetry-breaking transition from HLC to IHLC has not been observed in coupled {\it identical} oscillators. In this context it should be mentioned that this symmetry-breaking transition from HLC to IHLC should not be confused with the observation of Ref.~\cite{qstr2} where the authors observed a sudden transition from HLC to IHLC in a network of genetic oscillators under phase repulsive coupling: that transition resulted from the presence of multistability instead of symmetry-breaking in a limit cycle.

In this paper, we indeed observe the symmetry-breaking transition from HLC to IHLC in coupled {\em identical} paradigmatic oscillators under mean-field coupling with an additional filter in the self-feedback path. We identify that the IHLC-HLC transition arises due to the interplay of mean-field coupling and the local filtering. In earlier studies it has been established that the mean-field coupling can induce AD, OD, and AD-OD transition even in identical coupled oscillators \cite{tanpre1,tanpre2}. The mean-field coupling is very much relevant in biology and physics, e.g., in the context of genetic oscillators the diffusion of autoinducer molecules through the cell membrane is governed by the mean-field coupling with quorum-sensing mechanism \cite{qstr,qstr2}.  On the other hand in practical coupling path a signal may suffer dispersion and attenuation due to the change in phase and amplitude of the signal, respectively. If a signal suffers both dispersion and attenuation the coupling path (or channel) is said to act as a low-pass filter (LPF); whereas, if only dispersion occurs without any change in amplitude (i.e., the case of zero attenuation) the channel may be modeled as an all-pass filter (APF). LPFs are omnipresent in electrical and biological networks. Examples include: the  musculoskeletal system of human body has an inbuilt local low-pass filtering system \cite{skeleton-lpf}, abdominal ganglion of the crayfish contains LPFs \cite{cray-lpf}, a LPF is an essential building block of phase-locked loops \cite{chaos-dpll}. On the other hand, APFs have wide applications in electronic communication systems as active phase shifters \cite{sedra}. In hyperchaotic time-delayed systems the application of APF as time-delay block has recently been established \cite{book}. In biological and electrical  networks where time-delay or phase shift occur without any attenuation, the notion of all-pass filtering is very much relevant. For example, in neuronal systems, action potential propagates without any attenuation due to the perfect balance created by ion pumps and protein channels \cite{iz-book}; in electronic communication systems, hubs or local amplifiers are used to preserve the signal amplitude; However, in those cases, the signal invariably experiences a time delay or phase shift. 

The effect of a LPF has already been studied in the context of synchronization \cite{raj-lpf,laser-lpf}. Recently, \citet{zou-lpf} established that the presence of a LPF in the self-feedback path provides a general mechanism for rhythmogenesis, which is an important phenomenon as the cessation of oscillation often leads to a fatal system degradation and an irrecoverable malfunctioning in many physical, biological, and physiological systems  \cite{zou-prl13,kurthnat15,tanrhyth}. They considered diffusive coupling and show that depending upon time-delay or parameter mismatch the cut-off frequency of the LPF can control rhythmogenesis. However, a detailed bifurcation analysis is required in order to understand the exact genesis of rhythmogenesis. On the other hand, hitherto the effect of an all-pass filter on the dynamics of coupled oscillators has not been studied.   

In this paper we consider the effect of both low-pass and all-pass filtering and show that the IHLC-HLC transition is the consequence of the local filtering. With a rigorous bifurcation analysis we show that depending upon the interplay of filter and coupling parameters, the system at first goes through an AD-OD transition with increasing coupling strength and then the inhomogeneous stable steady state branches of the OD state become unstable through supercritical Hopf bifurcation giving rise to {\em stable} IHLC; this stable IHLC then experiences a pitchfork bifurcation of the limit cycle (PBLC) and gives rise to a HLC. Conversely, if one starts from a large coupling strength a HLC continuously transforms into IHLC through a PBLC. We also experimentally demonstrate the IHLC-HLC transition using van der Pol oscillators that proves the robustness of the transition scenario.

\section{Effect of local filtering in Stuart-Landau oscillators}
\label{sec2}
\subsection{Low-pass filter}
\label{sec:lpf}
We consider two Stuart-Landau oscillators interacting through mean-field diffusive coupling with local low-pass filtering. The mathematical model of the coupled system is given by
\begin{subequations}\label{ls}
\begin{align}
\dot{Z_j}&=(1+i\omega_j-|Z_j|^{2})Z_j+\epsilon\left(Q\overline{Z}-S_j\right),\\
\dot{S}_{j} &= \alpha(-S_{j}+{\mbox Re}(Z_{j})).\label{lpf}
\end{align}
\end{subequations}
with $j=1,2$; $\overline{Z}=\frac{1}{2}\sum_{j=1}^{2}{\mbox Re}(Z_j)$ is the mean-field of the coupled system, $Z_j=x_j+iy_j$. The individual Stuart-Landau oscillators have unit amplitude and eigenfrequency $\omega_j$ (in the rest of the paper we consider $\omega_j=\omega$, i.e., oscillators have the same eigenfrequency). $\epsilon$ represents the coupling strength, and $Q$ controls the density of mean-field \cite{qstr,srimali,tanchaosad,bandutta}; $ 0\leqslant Q \leqslant 1$. Equation \eqref{lpf} governs the dynamics of a LPF whose input is ${\mbox Re}(Z_{j})$: here $S_j$ represents the output of the LPF and $\alpha$ is the cut-off frequency or corner frequency. The limit $\alpha\rightarrow\infty$ represents the unfiltered case as then $S_{j}={\mbox Re}(Z_{j})$; smaller $\alpha$ imposes a stronger filtering effect because then higher frequencies and their harmonics get strongly attenuated. 

\begin{figure*}
\includegraphics[width=.80\textwidth]{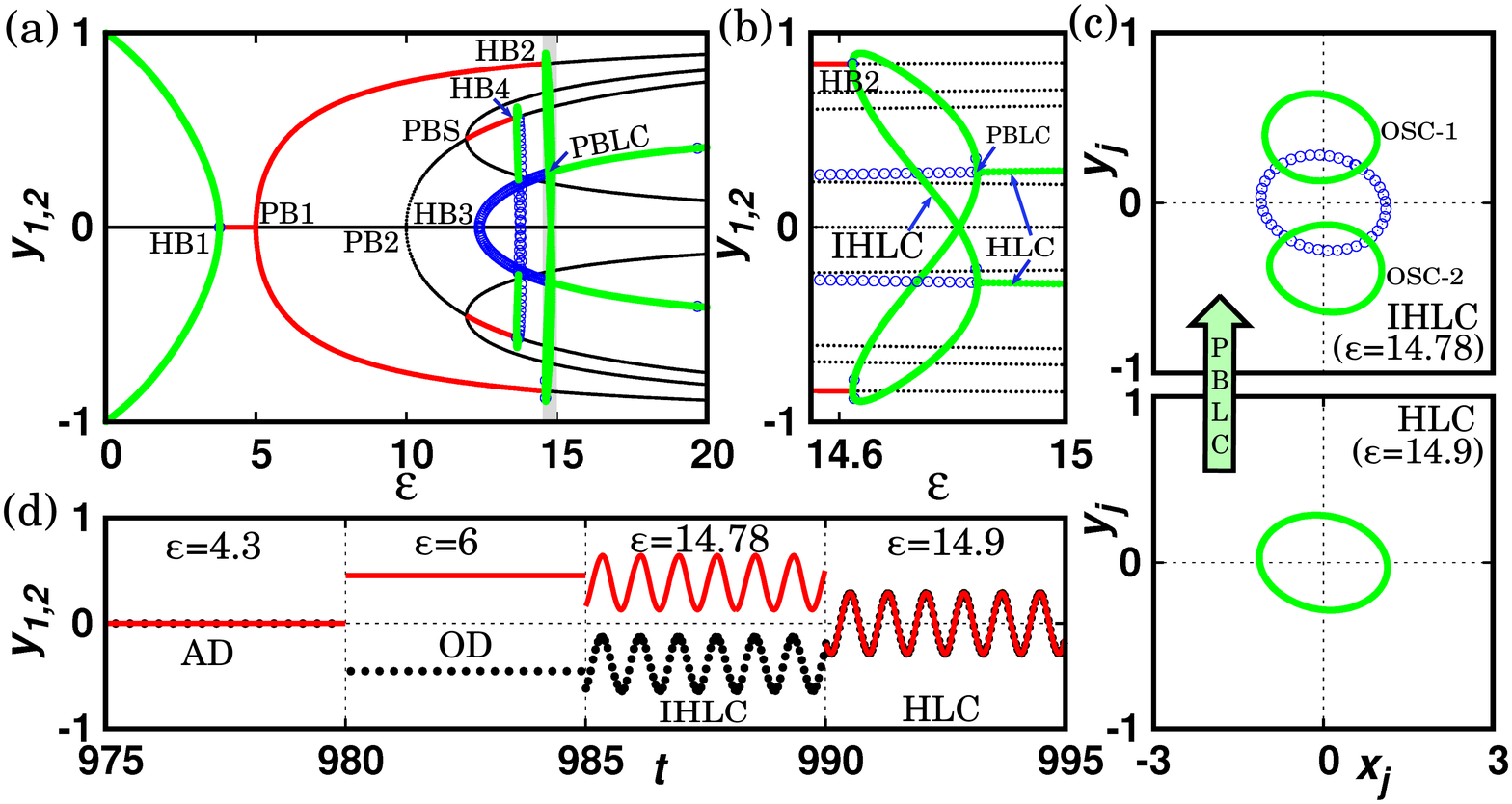}
\caption{(Color online) (a) Bifurcation diagram with $\epsilon$ (using XPPAUT) for $Q=0.5$, $\alpha=8$, and $\omega=2$ (Eq.~\ref{ls}). Grey (red) lines: stable fixed points, black lines: unstable fixed points, solid circle (green): stable limit cycle, open circle (blue): unstable limit cycle. HB\{1,2,3,4\} and PB\{1,2,S\} are Hopf and pitchfork bifurcation points, respectively. (b) Zoomed in view of the transition from stable IHLC to HLC; PBLC denotes the pitchfork bifurcation of limit cycle. (c) Phase space representation of the symmetry-breaking in limit cycle: Transition from {\it stable} HLC (lower panel, $\epsilon=14.78$) to {\it stable} IHLC (upper panel, $\epsilon=14.9$) through PBLC. Upper panel: Oscillator with $j=1$ ($j=2$) is denoted as OSC-1 (OSC-2) and unstable limit cycle is shown in open circle (blue). (d) Time series of $y_1$ (solid line) and $y_2$ (dotted line) showing AD ($\epsilon=4.3$), OD ($\epsilon=6$), IHLC ($\epsilon=14.78$), and HLC ($\epsilon=14.9$). Note that, the IHLCs are in phase with each other.}
\label{fig1}
\end{figure*}


Equation\eqref{ls} has the following fixed points: the origin $(0, 0, 0, 0, 0, 0)$ as the trivial fixed point, and additionally two coupling-dependent nontrivial fixed points: (i) Inhomogeneous steady state (IHSS) $\mathcal{F}_{IHSS}\equiv$ (${x}^\ast$, ${y}^\ast$, $-{x}^\ast$, $-{y}^\ast$, ${x}^\ast$, $-{x}^\ast$), where ${x}^\ast = -\frac{\omega {y}^\ast}{{\omega}^2 + \epsilon {{y}^\ast}^2},$ and $
{y}^\ast = \pm \sqrt {\frac{(\epsilon - 2{\omega}^2) + \sqrt{{\epsilon}^2 - 4{\omega}^2}}{2\epsilon}}.$ (ii) Nontrivial homogeneous steady state $\mathcal{F}_{NHSS}\equiv$ (${x}^\dagger$, ${y}^\dagger$, ${x}^\dagger$, ${y}^\dagger$, ${x}^\dagger$, ${x}^\dagger$), where ${x}^\dagger = - \frac{\omega {y}^\dagger}{\epsilon(1 - Q){{y}^\dagger}^2 + {\omega}^2},$ and ${y}^\dagger = \pm \sqrt{\frac{\epsilon(1 - Q) - 2 {\omega}^2 + \sqrt{{(\epsilon - \epsilon Q)}^2 - 4{\omega}^2}}{2\epsilon(1 - Q)}}$.

We can write the fixed points of the system as (${x}^m$, ${y}^m$, $P{x}^m$, $P{y}^m$, ${x}^m$, $P{x}^m$), where \{$x^m=0,~y^m=0$\} gives the trivial fixed point, \{$m=\ast$, $P=-1$\} represents the $\mathcal{F}_{IHSS}$ and \{$m=\dagger$, $P=1$\} gives $ \mathcal{F}_{NHSS}$. The Jacobian matrix of the system at the fixed point (${x}^m$, ${y}^m$, $P{x}^m$, $P{y}^m$, ${x}^m$, $P{x}^m$) is
\begin{equation}\label{jacobian}
{\mbox {\bf J}}=\left(\begin{array}{cccccc} A_{11} & A_{12}& \frac{\epsilon Q}{2} & 0 & B_{11} & 0 \\
A_{21}& A_{22} & 0 & 0 & 0 & 0\\
\frac{\epsilon Q}{2} & 0 &A_{11}& A_{12}& 0 & B_{11} \\
 0 & 0 & A_{21}& A_{22} & 0 & 0\\
\alpha & 0 & 0 & 0 & -\alpha & 0 \\
0 & 0 & \alpha & 0 & 0  & -\alpha \end{array}\right),
\end{equation}
where $A_{11}=\left(1-3{x^m}^2-{y^m}^2+\frac{\epsilon Q}{2}\right)$, $B_{11}=-\epsilon$, $A_{12}=\left(-2{x^m}{y^m}-\omega \right)$, $A_{21}=\left(\omega-2{x^m}{y^m} \right)$, and $A_{22}=(1-{x^m}^2-3{y^m}^2)$. Note that although $\alpha$ has no effect on the fixed points it affects their stability as the coefficients of the Jacobian matrix contains $\alpha$-dependent terms.

To derive the bifurcation points, we write the characteristic equation of the system at the fixed point (${x}^m$, ${y}^m$, $P{x}^m$, $P{y}^m$, ${x}^m$, $P{x}^m$) as
\begin{align}
\label{ce_revlp}
\left({\lambda}^3+P_2{\lambda}^2+P_1\lambda+P_0\right)\left({\lambda}^3+P'_2{\lambda}^2+P'_1\lambda+P'_0 \right)=0
\end{align}
where 
\begin{subequations}\label{Ps}
\begin{align}
P_2 &=-2+4{r^m}^2-\epsilon Q+\alpha,\\
P_1&=1+\alpha[4{r^m}^2-2+\epsilon(1-Q)]+{\omega}^2\\ \nonumber
&+3{r^m}^4-4{r^m}^2+\epsilon Q(1-{x^m}^2-3{y^m}^2),\\ 
P_0&=\alpha[1-4{r^m}^2+3{r^m}^4\\ \nonumber
&+{\omega}^2-\epsilon(1-Q)(1-{x^m}^2-3{y^m}^2)],\\
P'_2&=\alpha-2+4{r^m}^2,\\
P'_1&=\alpha[4{r^m}^2-2+\epsilon]+1+3{r^m}^4\\ \nonumber
&-4{r^m}^2+{\omega}^2,\\
P'_0&=\alpha[1-4{r^m}^2+3{r^m}^4+{\omega}^2\\ \nonumber
&-\epsilon(1-{x^m}^2-3{y^m}^2)],
\end{align}
\end{subequations} 
where ${r^m}^2=({x^m}^2+{y^m}^2)$. From the close inspection of the fixed points one can find two pitchfork bifurcations (PB) given by PB1 and PB2 occurring at
\begin{equation}
\label{pb}
{\epsilon}_{PB1}= 1+{\omega}^2,\quad {\epsilon}_{PB2} = \frac{1+{\omega}^2}{1-Q}.
\end{equation}
IHSS emerges at ${\epsilon}_{PB1}$. PB2 gives rise to a nontrivial HSS state. These two results are the same as those from the mean-field coupled Stuart-Landau oscillators (without filtering) \cite{tanpre1}. It is noteworthy that the occurrence of PB1 and PB2 do not depend upon $\alpha$, rather, as  we will see later that $\alpha$ controls their stability.


Before we proceed further with the stability analysis let us look at the bifurcation scenario (using XPPAUT \cite{xpp}) with a representative value of $Q=0.5$ and $\alpha=8$. Without any loss of generality in this paper we consider $\omega=2$. Fig.~\ref{fig1} (a) shows that, with increasing $\epsilon$, the coupled system experiences an inverse Hopf bifurcation at ${\epsilon}_{HB1}$ and an AD state emerges. With further increase in $\epsilon$, this AD state transforms into an OD state through a pitchfork bifurcation at ${\epsilon}_{PB1}$. 
The inhomogeneous steady state branches of the OD state loss their stability through the Hopf bifurcation (HB2) at ${\epsilon}_{HB2}$ and give rise to two {\it stable} inhomogeneous limit cycles (IHLC). This is in sharp contrast to the mean-field coupled oscillators of Ref.~\cite{tanpre1} where the OD branches, once created, remain stable for increasing coupling strength. Also, an additional Hopf bifurcation (HB3) of the trivial steady state emerges and gives birth of an unstable limit cycle. 
Interestingly, the stable IHLCs from HB2 collide with the unstable HLC created from HB3 and this collision creates a stable HLC through a pitchfork bifurcation of the limit cycle (PBLC) (see Fig.~\ref{fig1}(b) for a zoomed-in view). 
Here, the role of $\alpha$ is two-fold: it makes the OD branches unstable and perfectly organizes the location of HB2 and HB3 such that they govern the PBLC that creates the transition from IHLC to HLC. 

This transition can be visualized more clearly for a decreasing $\epsilon$. Figure~\ref{fig1}(c) shows the phase-space plot of the symmetry-breaking bifurcation of limit cycle: for $\epsilon>\epsilon_{PBLC}$ one has a stable HLC (lower panel, $\epsilon=14.9$). Now if we decrease $\epsilon$, the HLC experiences a PBLC and gives rise to two IHLCs (shown in green solid circles in the upper panel, $\epsilon=14.78$) and the HLC itself becomes unstable (shown in blue open circle). 
The IHLCs are then transformed into OD through HB2 and the unstable HLC disappears at HB3 (see Fig.~\ref{fig1}(a)). Therefore, HB2 and HB3 act as perfect hosts for the IHLCs and the unstable HLC, respectively. 
Another interesting limit cycle oscillation emerges through Hopf bifurcation (HB4) from the nontrivial HSS (NHSS) branches (created by a subcritical pitchfork bifurcation); see Fig.~\ref{fig1} (a). This is a bistable LC as depending upon the initial conditions both oscillators either oscillate in the upper or lower branch. Therefore, this LC can be denoted as a nontrivial homogeneous limit cycle (NT-HLC). However, we find that a slight asymmetry in the coupled systems (e.g., parameter mismatch) causes the NT-HLC to vanish. 
Finally, the time series of $y_{1,2}$ at different dynamical states with representative values of $\epsilon$ are shown in Fig.~\ref{fig1}(d) (with $Q=0.5$ and $\alpha=8$): apart from AD and OD it shows the IHLCs, i.e., limit cycles with shifted origin (for $\epsilon=14.78$) and HLC, i.e., limit cycles around zero origin (for $\epsilon=14.9$). Note that, the IHLCs are in phase with each other, which is expected as the oscillators are identical and coupled under a symmetric mean-field coupling.

To understand the role of $\alpha$ quantitatively, we derive the important bifurcation curves using the characteristic equation \eqref{ce_revlp}. Since \eqref{ce_revlp} is a sixth-order polynomial it is difficult to extract bifurcation points from the eigenvalue analysis. Therefore, we use the technique used in \cite{math}, where it has been shown that one can predict the Hopf bifurcation points from the {\it coefficients} of the characteristic equation itself. From \eqref{ce_revlp} the analytical expressions of HB1 and HB3 are obtained by putting $|P_1P_2-P_0|_{(x^m=0, y^m=0)}=0$ (note that HB1 and HB3 are the bifurcation points associated with the trivial fixed point $x^m=0, y^m=0$). From Eq.~\ref{Ps}, using $P_2=(-2-\epsilon Q+\alpha)$, $P_1=\alpha\{-2+\epsilon(1-Q)\}+1+\epsilon Q+{\omega}^2$, $P_0=\alpha\{1+{\omega}^2-\epsilon(1-Q)\}$ we derive
\begin{subequations}
\begin{align}
{\alpha}_{HB1} &= \frac{-B_{HB3}-\sqrt{{B_{HB3}}^2-4A_{HB3}C_{HB3}}}{2A_{HB3}},\\
{\alpha}_{HB3} &= \frac{-B_{HB3}+\sqrt{{B_{HB3}}^2-4A_{HB3}C_{HB3}}}{2A_{HB3}},\label{hb3}
\end{align}
\end{subequations}
where $A_{HB3}=\epsilon-(\epsilon Q+2)$, $B_{HB3}=(\epsilon Q+2)^2-\epsilon(\epsilon Q +1)$, $C_{HB3}=-(\epsilon Q+2)(1+{\omega}^2+\epsilon Q)$.

Since HB2 is associated with the IHSS branch of OD, its locus is obtained by using $|P_1P_2-P_0|_{\mathcal{F}_{IHSS}}=0$, yielding 
\begin{equation}\label{hb2}
{\alpha_{HB2}}= \frac{-B_{HB2}+\sqrt{{B_{HB2}}^2-4A_{HB2}C_{HB2}}}{2A_{HB2}},
\end{equation}
where 
\begin{align*}
A_{HB2}&=2+\epsilon(1-Q)-\frac{8{\omega}^2}{L_{HB2}},\\ \nonumber
B_{HB2}&={A_{HB2}}^2+{\epsilon}^2Q+(2{\omega}^2-{\epsilon}^2-4\epsilon)+\frac{{\omega}^2(4+10\epsilon)}{L_{HB2}},\\ \nonumber
C_{HB2}&=\left[{\omega}^2(1+2Q)-QL_{HB2}+\frac{12{\omega}^4}{L_{HB2}^2}+\frac{2{\omega}^2(\epsilon Q-2)}{L_{HB2}}\right]\\ \nonumber
&(A_{HB2}-\epsilon),\\ \nonumber
L_{HB2}&=(\epsilon+\sqrt{{\epsilon}^2-4{\omega}^2}). \nonumber
\end{align*}
Similarly, the locus of HB4 is obtained by using $|P_1P_2-P_0|_{\mathcal{F}_{NHSS}}=0$, which gives
\begin{equation}\label{hb4}
{\alpha_{HB4}}= \frac{-B_{HB4}+\sqrt{{B_{HB4}}^2-4A_{HB4}C_{HB4}}}{2A_{HB4}},
\end{equation}
where $A_{HB4}=2+\epsilon(1-Q)-\frac{8{\omega}^2}{L_{HB4}}$, $B_{HB4}={A_{HB4}}^2-{\epsilon}^2(1-Q)-4\epsilon+\frac{2{\omega}^2}{(1-Q)}+\frac{{\omega}^2\{4+10\epsilon(1-Q)\}}{(1-Q)L_{HB4}}$, $C_{HB4}=\left[\frac{{\omega}^2(1+Q)}{(1-Q)}-\frac{QL_{HB4}}{(1-Q)}+\frac{12{\omega}^4}{{L_{HB4}}^2}+\frac{2{\omega}^2(\epsilon Q-2)}{L_{HB4}}\right](A_{HB4}-\epsilon)$, $L_{HB4}=[\epsilon(1-Q)+\sqrt{{\epsilon}^2(1-Q)^2-4{\omega}^2}]$.

Figure~\ref{fig2}(a) shows the two-parameter bifurcation curves in the $\epsilon-\alpha$ space for $Q=0.5$ (using XPPAUT and the analytical results obtained in Eq.~\ref{pb}--Eq.~\ref{hb4}). It demonstrates that the zone of the death region (determined by the HB1 and HB2 curves) decreases with decreasing $\alpha$. The OD state loses its stability through HB2 with an increasing $\epsilon$. In the absence of filtering $\alpha\rightarrow \infty$ and $\epsilon_{HB2}\rightarrow \infty$, therefore, in the unfiltered case the OD state never losses its stability with increasing $\epsilon$: as a consequence, neither rhythmogenesis nor IHLC-HLC transition occur without filtering. As we decrease the cut-off frequency $\alpha$, the OD state losses stability through Hopf bifurcation (HB2) for lower values of $\epsilon_{HB2}$. Let us investigate the subtlety of the IHLC-HLC transition scenario in more detail. Based upon the value of $\alpha$, we identify four distinct dynamical regions:

\begin{figure}
\includegraphics[width=.5\textwidth]{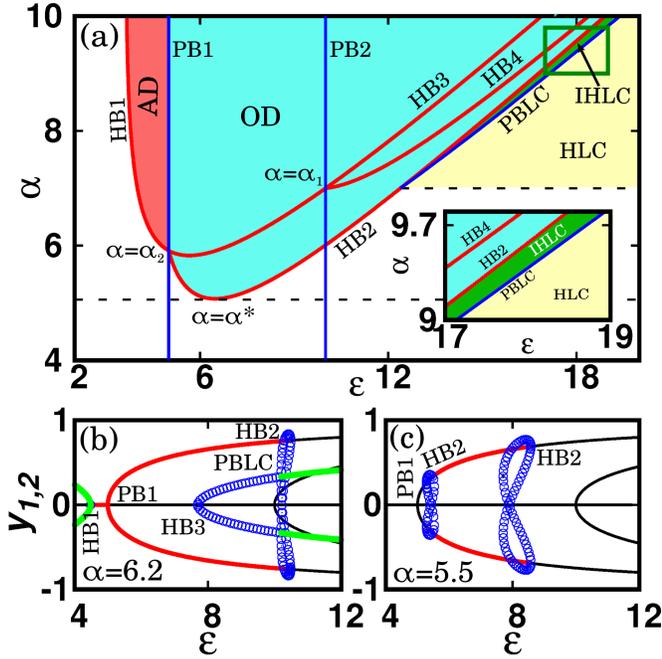}
\caption{\label{fig2} (Color online) (a) Low-pass filtering: Two parameter bifurcation diagram for Stuart-Landau oscillators of \eqref{ls} ($Q=0.5$). IHLC occurs in the dark gray (green) shaded region bounded by HB2 and PBLC curve. Yellow shaded region is for $\alpha\ge\alpha_1$. Inset: a zoomed-in view of the IHLC-HLC transition. White or yellow shaded region is for limit cycle. (b) Single parameter bifurcation for $\alpha=6.2$ (i.e., $\alpha_2<\alpha<\alpha{_1}$). Note that IHLC now becomes unstable although HLC is stable. (c) Bifurcation for $\alpha=5.5$ (i.e., $\alpha^*<\alpha<\alpha{_2}$): a single OD region exists between two HB2. Here $\omega=2$.}
\end{figure}

(i) $\alpha>\alpha{_1}$, IHLC-HLC transition: In this region the transition from IHLC to HLC occurs. Here $\alpha_1$ is the value of $\alpha$ for which HB3 and HB4 collide with PB2. This can be derived from \eqref{hb3}, \eqref{hb4}, and \eqref{pb} as
\begin{equation}\label{alpha1}
\alpha_1=\frac{-B_\alpha\pm \sqrt{{B_\alpha}^2-4A_\alpha C_\alpha}}{2A_\alpha},
\end{equation}
where, $A_\alpha=(1-{\omega}^2)(Q-1)^2$, $B_\alpha=({\omega}^2-3)({\omega}^2Q+1)-Q^2({\omega}^2-1)^2+4Q$ and $C_\alpha=({\omega}^2+1)[({\omega}^2-1)Q+2]$. Note that although the IHLC-HLC transition is the result of the bifurcation of the limit cycle, however, its stability is controlled by the bifurcation of the fixed points, i.e., HB3, HB4 and PB2. In Fig.~\ref{fig2}(a), the zone of IHLC is bounded in between the HB2 and PBLC curves (shown with dark gray (green) shading; see also the inset); HLC appears below the PBLC curve. For $\alpha>\alpha{_1}$, with the variation of $\epsilon$, the IHLC-HLC transition is bounded by the PBLC curve and the line of $\alpha=\alpha_1$ (here $\alpha_1=7$ using \eqref{alpha1} for $Q=0.5$ and $\omega=2$): this zone is shown in yellow shading for visual guidance. Figure~\ref{fig1}(a) shows the bifurcation diagram in this region for a representative value $\alpha=8$. 

(ii) $\alpha_2<\alpha<\alpha{_1}$, unstable IHLC, stable HLC: At $\alpha=\alpha_2$, HB3 collides with PB1. Its value can be derived from \eqref{hb3} and \eqref{pb} (expression not shown here). In this region the IHLC becomes unstable and the Hopf bifurcation HB2 is now a subcritical one. However, the unstable LC originated from HB3 still collides with the unstable LC emanated from HB2 and gives a stable HLC for $\epsilon>\epsilon_{PBLC}$. Therefore, in this (narrow) region of $\alpha$ we do not have the (stable) IHLC-HLC transition; Fig.~\ref{fig2}(b) shows this scenario for $\alpha=6.2$. 

(iii) $\alpha^*<\alpha<\alpha{_2}$: $\alpha^*$ represents the minima of the HB2 curve (see Fig.~\ref{fig2}(a)), which can be derived by minimizing \eqref{hb2}. Here the HB2 curve becomes multivalued for a single $\alpha$. Fig.~\ref{fig2}(c) shows the representative bifurcation diagram in this region (for $\alpha=5.5$). Here no AD-OD transition is possible, instead a solitary OD region is interspersed in the limit cycles.

(iv) $\alpha<\alpha^*$: Complete rhythmogenesis, i.e., the system enters into oscillatory state for any coupling strength.   

\begin{figure}
\includegraphics[width=.48\textwidth]{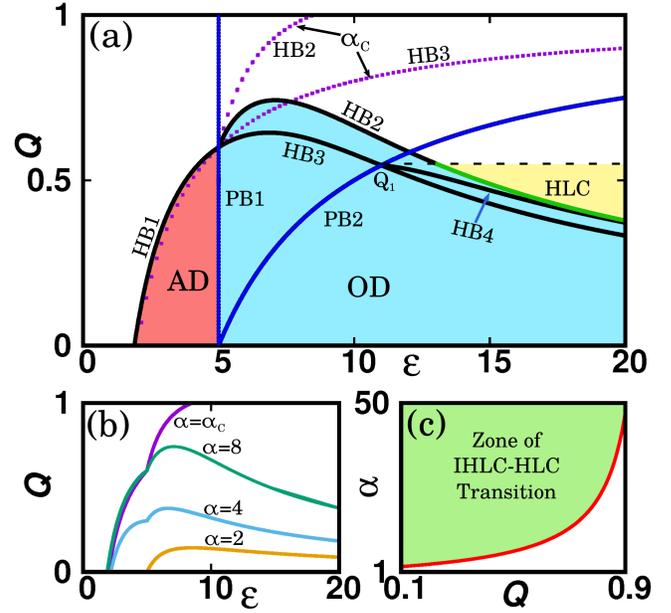}
\caption{\label{fig3} (Color online) Stuart-Landau oscillator with LPF. (a) Two-parameter bifurcation diagram in the $\epsilon-Q$ space for an unfiltered case ($\alpha=\alpha_c\rightarrow\infty$) and LPF with $\alpha=8$. For $Q<Q_1$ IHLC to HLC transition occurs near the HB2 curve (highlighted in green color). White or yellow shaded region is for limit cycle. (b) Effect of $\alpha$ on the rhythmogenesis: Shown are the critical curves for $\alpha=\alpha_c$ (i.e., unfiltered case), $\alpha=8,4,2$. Area under the curve represents the death region: the death region is quenched with decreasing $\alpha$. (c) The curve in the $Q-\alpha$ space showing the zone of the IHLC-HLC transition. Here $\omega=2$.}
\end{figure}

Next, we explore the effect of $\alpha$ in the $\epsilon-Q$ parameter space. Figure~\ref{fig3}(a) shows this for two cases: one is without any local filtering, i.e., $\alpha=\alpha_c\rightarrow\infty$ and the other is shown with local LPF for $\alpha=8$. Note the effect of $\alpha$ in order to create an oscillation from the OD branch: it actually bends the HB2 and HB3 curves of the conventional (i.e., unfiltered) case downwards to create an oscillation and therefore induces rhythmogenesis with increasing $\epsilon$. We get the IHLC-HLC transition for $Q<Q_1$, where $Q_1$ is the value where the HB3 and HB4 curves collide with the PB2 curve. The rhythmogenesis is actually facilitated by decreasing $\alpha$. Figure~\ref{fig3}(b) shows that the zone of the death region is quenched with decreasing $\alpha$. The parameter zone of observing the IHLC-HLC transition is shown in Fig.~\ref{fig3}(c) in the $Q-\alpha$ parameter space (using Eq.~\ref{alpha1}): from this we can prescribe the condition for observing HLC-IHLC transition -- vary $\epsilon$ with $\alpha>\alpha_1$ and $Q<Q_1$ (i.e., the upper part of the $Q-\alpha$ curve). 

\subsection{All-pass filter}\label{apf:sub}
Next, we investigate the effect of an all-pass filter in the local feedback path. The mathematical model of the coupled Stuart-Landau oscillators under mean-field coupling and an all-pass filter is then given by
\begin{subequations}\label{apf}
\begin{align}
\dot{Z_j}&=(1+i\omega_j-|Z_j|^{2})Z_j+\epsilon\left(Q\overline{Z}-U_j\right),\\
\dot{S}_{j} &= \alpha(-S_{j}+{\mbox Re}(Z_{j})),\\
U_j&=2S_j-{\mbox Re}(Z_{j}).
\end{align}
\end{subequations}
Equation \eqref{apf} is a differential-algebraic equation that governs the dynamics of an all-pass filter (APF) whose input is ${\mbox Re}(Z_{j})$: here $U_j$ is the output of the APF. In this case also $\alpha$ has the same meaning as \eqref{lpf}, however, it has a different effect on $U_j$: $\alpha$ does not change the amplitude of $U_j$ but it only controls the phase part (see Appendix~\ref{app}). Eq.~\eqref{apf} has the same set of fixed points as Eq.~\eqref{ls}, however, the Jacobian matrix of \eqref{apf} is modified from \eqref{jacobian} as now the elements $A_{11}$ and $B_{11}$ become $A_{11}=\left(1-3{x^m}^2-{y^m}^2+\frac{\epsilon Q}{2}\right)+\epsilon$ and $B_{11}=-2\epsilon$; other elements remain the same. 

An analysis in line of the previous subsection reveals that all the steady state bifurcation points are the same as those of the LPF case given in \eqref{pb}. The two-parameter bifurcation structure in the $\epsilon-\alpha$ space is shown in Figure~\ref{f:apf}(a) (for $Q=0.5$) using XPPAUT and analytically obtained bifurcation curves (expressions are given in Appendix~\ref{app2}). It can be observed that the qualitative structure of the Hopf curves in the two parameter space remains the same as that of the LPF case (Fig.~\ref{fig2}(a)), except the fact that now the minima of the curve HB1 determines $\alpha^*$-- the value of $\alpha$ below which complete rhythmogenesis sets in. In this case  there also exists a value of $\alpha$ where the HB3 and HB4 curves collide with the PB2 line: for an $\alpha$ greater than this value (shown with horizontal dashed line) the system shows an IHLC-HLC transition. Figure~\ref{f:apf}(b) demonstrates the transition from IHLC to HLC for increasing $\epsilon$ for $\alpha=20$ ($Q=0.5$). Figure~\ref{f:apf}(c) shows the bifurcation for $\alpha=10.85$, which shows a solitary AD state interspersed in the limit cycle region, because here the HB1 curve becomes multivalued; this is in contrast to the LPF case, where we get a solitary OD region due to the multivalued HB2 curve (cf. Fig.~\ref{fig2}(c)).
\begin{figure}
\includegraphics[width=.5\textwidth]{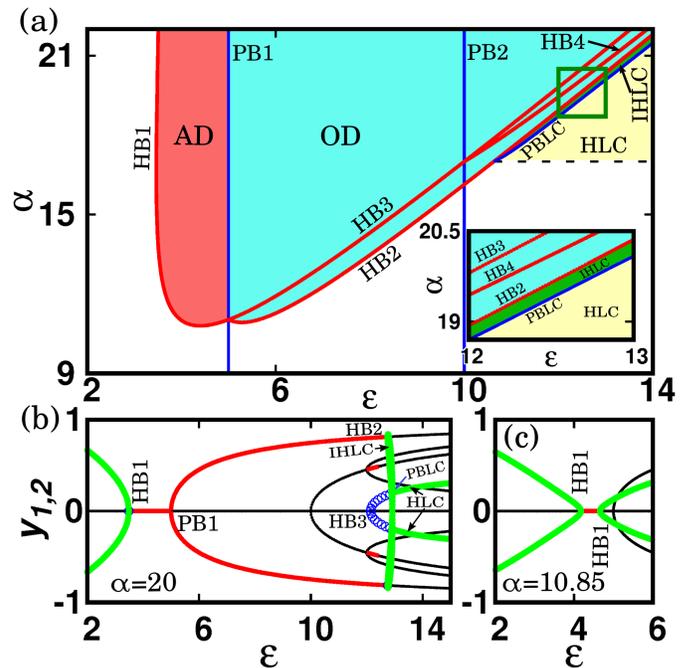}
\caption{\label{f:apf} (Color  online) (a) All-pass filtering: Two parameter bifurcation diagram of Stuart-Landau oscillators of \eqref{apf} in the $\epsilon-\alpha$ parameter space for $Q=0.5$. Inset: a zoomed-in view of the IHLC-HLC transition. White or yellow shaded region is for limit cycle. (b) Single parameter bifurcation for $\alpha=20$. Note the IHLC-HLC transition through PBLC. (c) Bifurcation for $\alpha=10.85$; a single AD region exists between two HB1 points. Other parameter: $\omega=2$.}
\end{figure}

Further, it is noteworthy that for the parameters same as in the LPF case, an APF can revoke the death states even for a comparatively higher value of $\alpha$. Note that a lesser $\alpha$ (i.e., a lesser cut-off frequency) means a stronger filtering effect. Therefore, even a weaker {\it all-pass filtering} is equivalent to a comparatively stronger {\it low-pass filtering} as far as the rhythmogenesis is concerned. This is due to the fact that for a given $\alpha$ the phase shift introduced by an APF is twice of that of a LPF (see Appendix \ref{app}).

\begin{figure}
\includegraphics[width=.5\textwidth]{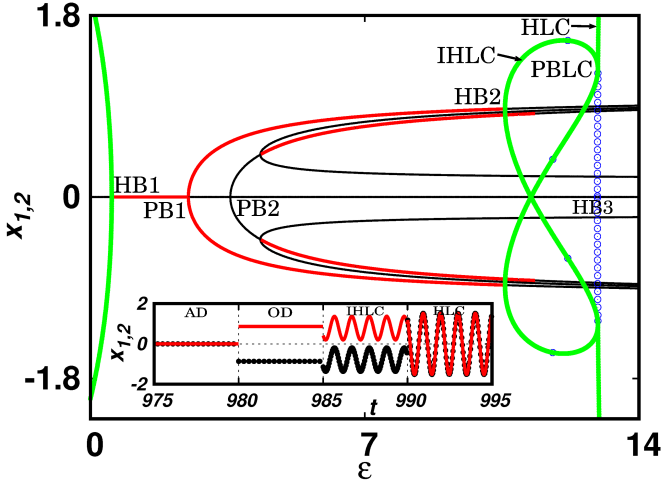}
\caption{\label{f:vdp} (Color online) Bifurcation diagram for van der Pol oscillators (Eq.~\ref{vdp}). $\alpha=4$, $Q=0.3$. Inset shows the transitions between the following regions: AD ($\epsilon=2$), OD ($\epsilon=10$), IHLC ($\epsilon=11$), and HLC ($\epsilon=12.96$). Here $a=0.4$.}
\end{figure}
\section{Effect of local filtering in van der Pol oscillators}
To verify the generality of the observed transitions, we consider two van der Pol  (vdP) oscillators interacting through mean-field diffusive coupling with local filtering; the mathematical model of the coupled system is given by
\begin{subequations}
\label{vdp}
\begin{align}
\dot{x}_{j} &= y_j+\epsilon\left(Q\overline{X}-F_{j}\right),\\
\dot{y}_{j} &= a(1-x_j{^2})y_j-x_j,\\
\dot{S}_{j}&=\alpha(-S_j+x_j).
\end{align}
\end{subequations}
Here $j=1,2$. $F_j$ represents the filtered local feedback term: for a LPF, $F_j=S_j$ and for an APF, $F_j=(2S_j-x_j)$. $\overline{X}=\frac{1}{2}\sum_{j=1}^{2}x_j$ is the mean-field term. The parameter $a$ determines the amplitude and shape of the oscillations. Equation~\eqref{vdp} has the following fixed points: the trivial fixed point is the origin $(0, 0, 0, 0, 0, 0)$ and two coupling dependent fixed points: (i) (${x}^\ast$, ${y}^\ast$, $-{x}^\ast$, $-{y}^\ast$, ${x}^\ast$, $-{x}^\ast$) where ${x}^\ast =\frac{{y}^\ast}{{\epsilon}}$ and ${y}^\ast = \sqrt {\epsilon^2-\frac{\epsilon}{a}}$. (ii) (${x}^\dagger$, ${y}^\dagger$, ${x}^\dagger$, ${y}^\dagger$, ${x}^\dagger$, ${x}^\dagger$) where ${x}^\dagger = \frac{{y}^\dagger}{\epsilon(1 - Q)}$ and ${y}^\dagger = \sqrt {\epsilon^2(1-Q)^2-\frac{\epsilon(1-Q)}{a}}$. The Jacobian matrix of the system at a fixed point is given by:
\begin{equation}\label{jacobian_vdp}
{\mbox {\bf J}}=\left(\begin{array}{cccccc} A_{11} & A_{12}& \frac{\epsilon Q}{2} & 0 & B_{11} & 0 \\
A_{21}& A_{22} & 0 & 0 & 0 & 0\\
\frac{\epsilon Q}{2} & 0 &A_{11}& A_{12}& 0 & B_{11} \\
 0 & 0 & A_{21}& A_{22} & 0 & 0\\
\alpha & 0 & 0 & 0 & -\alpha & 0 \\
0 & 0 & \alpha & 0 & 0  & -\alpha \end{array}\right),
\end{equation}
where $A_{12}=1$, $A_{21}=(-2ax^m y^m -1)$, $A_{22}=a(1-{x^m}^2)$ (we use the same sign convention as in \eqref{jacobian}). For a LPF, $A_{11}=\frac{\epsilon Q}{2}$ and $B_{11} = -\epsilon$ whereas for an APF, $A_{11}=\frac{\epsilon Q}{2}+\epsilon$ and $B_{11} = -2\epsilon$. 
One can derive the bifurcation points in line of the analysis of the previous section. Through a detailed bifurcation analysis, we find that in the case of vdP oscillators the bifurcation scenarios and the IHLC-HLC transition remain qualitatively the same as those of the Stuart-Landau oscillator. Figure~\ref{f:vdp} shows the representative bifurcation diagram with LPFs (for $\alpha=4$) (we take $a=0.4$ and $Q=0.3$); it shows the IHLC to HLC transition through PBLC. The inset of Fig.~\ref{f:vdp} demonstrates the time series depicting the AD, OD, IHLC and HLC for increasing $\epsilon$. For an APF we get the same transition scenario for a properly chosen value of $\alpha$ (results not shown here). 
\begin{figure}
\centering
\includegraphics[width=0.48\textwidth]{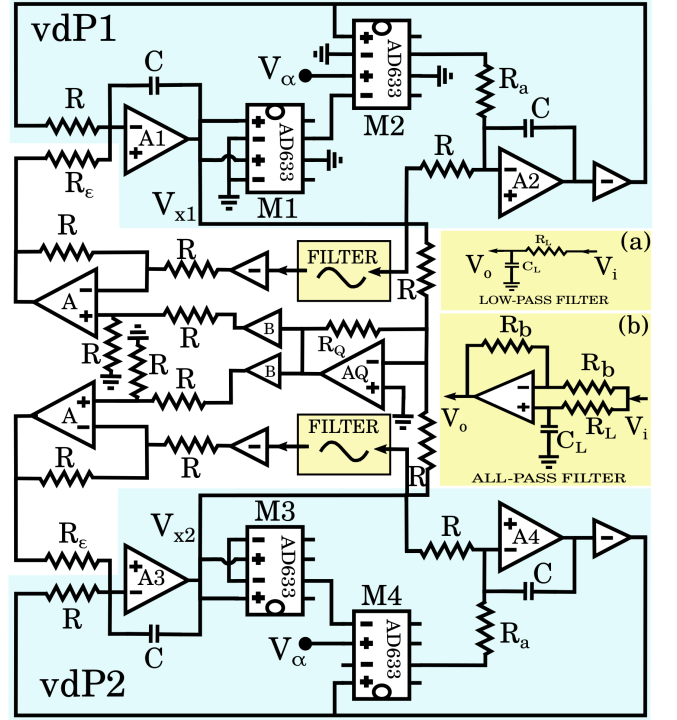}
\caption{Experimental electronic circuit diagram of coupled van der Pol (vdP) oscillators of Eq.~\ref{ckteq}. A1--A4, A and AQ are TL074 opamps. The buffers ``B'' are opamp based and inverters are realized using unity-gain inverting amplifiers. M1--M4 are multiplier chips (AD633JN). In the box labeled ``FILTER'' we use either LPF or APF, which are shown in insets (a) and (b), respectively. $R=10$ k\ohm, $R_a=286$ \ohm, $C=10$ nF, $C_L=0.1$ $\mu$F, $R_b=2.2$ k\ohm~ and $V_\alpha=0.1$ V. The resistors (capacitors) have $\pm 5\%$ ($\pm 1\%$) tolerance. We use a $\pm 15$ V power supply in the experiment.}
\label{ckt}
\end{figure}

\section{Experiment}
The coupled system of Eq.~\eqref{vdp} is implemented in an electronic circuit \cite{vdpckt}. The schematic of the circuit diagram is given in Fig.~\ref{ckt}. The individual van der Pol oscillators are shown in the shaded regions of the figure labeled ``vdP1'' and ``vdP2''. The vdP oscillators are coupled through the mean-field coupling scheme along with local filtering (low-pass or all-pass). We replace the box labeled ``FILTER'' by LPF or APF. The circuit of LPF and APF are shown in the inset of the figure. The outputs from the individual vdP oscillators are fed to an weighted inverting adder AQ which produces the mean-field given by $V_Q=-\frac{2R_Q}{R}\sum_{j=1}^2\frac{V_{xj}}{2}$. The coupling strength is controlled by the resistances $R_\epsilon$.

The voltage equation of the circuit of Fig.~\ref{ckt} can be written as
\begin{subequations}\label{ckteq}
\begin{align}
CR\dot{V}_{xj}&=V_{xj}+\frac{R}{R_\epsilon}\bigg[\frac{2R_Q}{R}\sum_{j=1}^2\frac{V_{xj}}{2}-F_{xj}\bigg],\label{ckteqa}\\
CR\dot{V}_{yj}&=\frac{R}{R_a}\bigg(V_\alpha-\frac{V_{xj}^2}{10}\bigg)\frac{V_{yj}}{10}-V_{xj},\label{ckteqb}\\
CR\dot{S}_{xj}&=\frac{CR}{R_LC_L}(-S_{xj}+V_{xj}),\label{ckteqc}
\end{align}
\end{subequations}
where $j=1,2$. For a LPF, $F_{xj}={S}_{xj}$ and for an APF, $F_{xj}=(2{S}_{xj}-V_{xj})$. We consider the following identities to normalize Eq.~\ref{ckteq}: $\epsilon=\frac{R}{R_\epsilon}$, $Q=\frac{2R_Q}{R}$, $a=\frac{R}{100R_a}$, $10V_\alpha=1$, $\alpha=\frac{RC}{R_LC_L}$, $x_j=\frac{V_{xj}}{V_{sat}}$, $y_j=\frac{V_{yj}}{V_{sat}}$ and $S_j=\frac{S_{xj}}{V_{sat}}$. Here $V_{sat}$ is the saturation voltage of the opamp. With these quantities Eq.~\eqref{ckteq} becomes equivalent to Eq.~\eqref{vdp}. In the experiment we choose the following values: $V_\alpha=0.1$ V, $C=10$ nF, $R=10$ k\ohm, $R_a=253$ \ohm~(i.e., $a=0.4$). The values of $\epsilon$ and $Q$ are varied by changing the resistances $R_\epsilon$ and $R_Q$, respectively (using POTs). 

\begin{figure}
\centering
\includegraphics[width=0.47\textwidth]{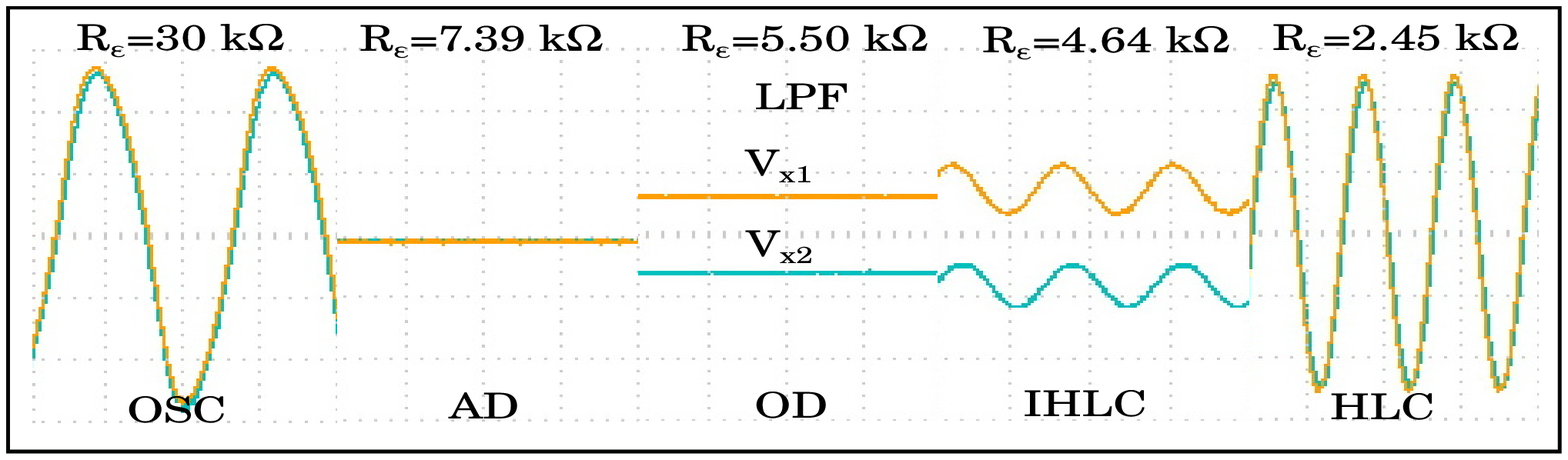}

\caption{(Color online) Low-pass Filter: Snapshots of experimental time series traces of $V_{x1}$ and $V_{x2}$. Synchronized LC (OSC) at $R_\epsilon=30$ k\ohm, AD at $R_\epsilon=7.39$ k\ohm, OD at $R_\epsilon=5.50$ k\ohm, IHLC at $R_\epsilon=4.64$ k\ohm~, and HLC at $R_\epsilon=2.45$ k\ohm. $R_L=862$ \ohm~and $R_Q=1.19$ k\ohm. Scale: $x$-axis, 25$\mu$s; $y$-axis, 1.25 v/div. See text for other parameters.}
\label{expt-lpf}
\end{figure}

\begin{figure}
\centering
\includegraphics[width=0.47\textwidth]{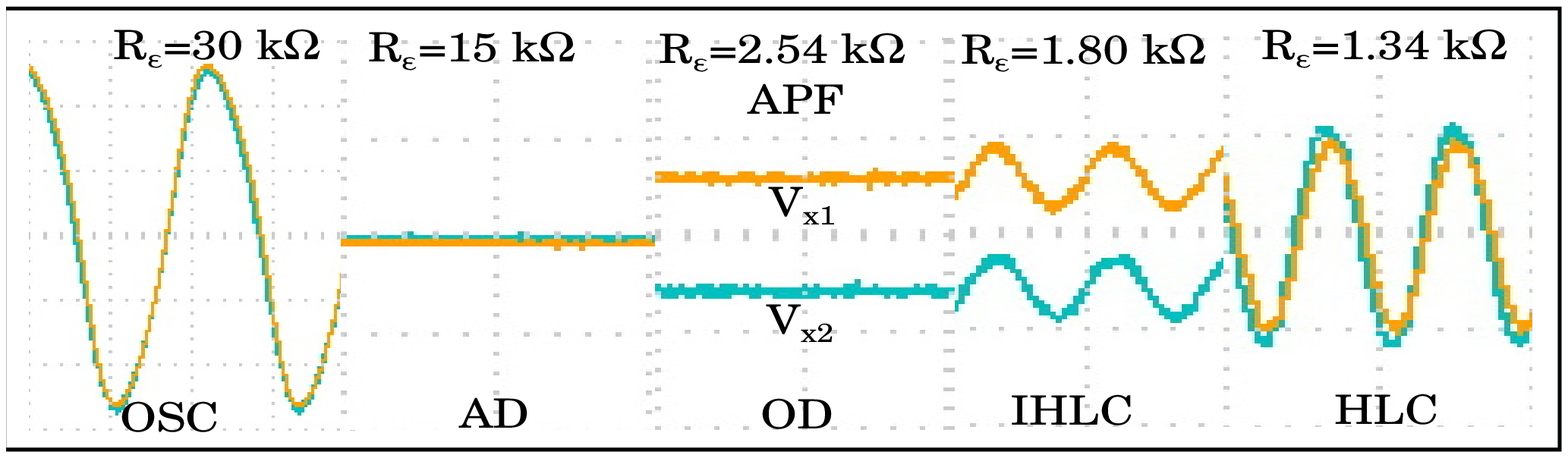}
\caption{(Color online) All-pass filter: Snapshots of experimental time series traces of $V_{x1}$ and $V_{x2}$. Synchronized LC at $R_\epsilon=30$ k\ohm, AD at $R_\epsilon=15$ k\ohm, OD at $R_\epsilon=2.54$ k\ohm, IHLC at $R_\epsilon=1.80$ k\ohm~, and HLC at $R_\epsilon=1.34$ k\ohm. $R_L=119$ \ohm~and $R_Q=1.134$ k\ohm. Scale: OSC panel same as Fig.~\ref{expt-lpf}; other panels: $x$-axis, 10$\mu$s; $y$-axis, 2.5 v/div. See text for other parameters.}
\label{expt-apf}
\end{figure}

In an experiment with LPF, we take $R_L=862$ \ohm~and $R_Q=1.19$ k\ohm~and decrease $R_\epsilon$ (increase $\epsilon$). The results of the experiment are summarized in Fig.~\ref{expt-lpf} with the snapshots of time series (taken using a digital storage oscilloscope, Tektronix TDS2002B, $60$ MHz, 1 GS/s). The same with an APF is shown in Fig.~\ref{expt-apf} for $R_L=119$ \ohm~and $R_Q=1.134$ k\ohm. 
In both cases we observe the following general scenario: with decreasing $R_\epsilon$ (i.e., increasing $\epsilon$) the system enters into the OD state from the synchronized oscillatory state via the AD state. A further  decrease in $R_\epsilon$ makes the system to oscillate around the inhomogeneous steady states (IHSS) giving IHLC oscillation. 
This IHLC is transformed into a HLC for further decrease in $R_\epsilon$. Therefore, with proper values of the filter parameter (here $\alpha$), despite the presence of fluctuation, noise and inherent parameter mismatch in the real experimental set up, we indeed observe a transition from IHLC to HLC which establishes that this transition is robust.

\section{Conclusion}
In this paper, we have discovered that the presence of a local filtering in the coupling path gives birth to an interesting transition from homogeneous limit cycle to inhomogeneous limit cycle in mean-field coupled identical oscillators. Using a rigorous bifurcation analysis we have explored the genesis of this transition. 
Unlike \cite{zou-lpf}, which studied the effect of low-pass filtering in the context of rhythmogenesis, here we have considered both low-pass and all-pass filtering and unravel the rich bifurcation structure associated with the coupled identical oscillators. 
Further, we have proposed a differential-algebraic equation to model an APF and for the first time explored the effect of APFs on the collective behavior of coupled oscillators.
We have also provided the first experimental evidence of the filtering induced rhythmogenesis and the IHLC-HLC transition.

The role of filtering (i.e., $\alpha$) on the rhythmogenesis and the IHLC-HLC transition can also be understood from dynamical point of view: the cut-off (or corner) frequency, $\alpha$, actually controls the rate of dissipation in the coupling term by controlling either the amplitude and phase (for a LPF) or only the phase (for an APF) of the self-feedback signal. A smaller $\alpha$ means lesser dissipation, which is conducive to rhythmogenesis and, therefore, to the reported transition.

Also, our study suggests that, as far as rhythmogenesis is concerned, APFs are more efficient than LPFs for the same parameter value (here $\alpha$). This is owing to the fact that for a given $\alpha$ an APF introduces more phase shift than that of a LPF (Appendix \ref{app}). It also suggests that the frequency selectivity of phase (instead of amplitude) is sufficient to induce rhythmogenesis and the observed IHLC-HLC transition. 

The next natural extension of this work will be to study the reported transition in networks of natural oscillators under diverse coupling schemes. We strongly believe that it will unravel the connection among several symmetry-breaking states, such as oscillation death, inhomogeneous limit cycles and amplitude chimeras. 

\appendix
\section{All-Pass Filter: Electronic analog}\label{app}
The electronic circuit of an all-pass filter (APF) is shown in Fig.~\ref{ckt}(b). Using Kirchhoff's voltage law, the voltage equation of the $R_L-C_L$ part reads 
\begin{equation}\label{a1}
\frac{dV}{dt}=\frac{1}{C_L R_L}(-V+V_i),
\end{equation}
where $V_i$ is the input voltage and $V$ is the voltage across the capacitor $C_L$. The parameter $(\frac{1}{C_L R_L})$ is the corner frequency $\alpha$. Again, from the opamp equation one gets the output of the APF as
\begin{equation}\label{Beta}
V_{0}=2V-V_i.
\end{equation}
Equation \eqref{a1} along with \eqref{Beta} represent the  differential-algebraic dynamical equation of an APF.

To show that the circuit of Fig.~\ref{ckt}(b) indeed represents an APF, we derive the frequency domain transfer function of the circuit as:
\begin{equation}\label{Gamma}
\frac{V_0}{V_i}=\frac{1-i\omega C_L R_L}{1+i\omega C_L R_L}=A_0 \exp(-i2\theta),
\end{equation} 
where $A_0=1$ and $\theta=\tan^{-1}(\omega C_L R_L)$. Note that $A_0=1$ ensures that the amplitude is frequency independent (unlike a LPF). The output only experiences a frequency dependent phase shift of $\phi=2\theta$. Also, it is interesting to note that for the same $\alpha$ (i.e., $\frac{1}{C_LR_L}$) the phase shift introduced by a LPF (i.e., $\theta$) is half of that of an APF (i.e., $\phi$).
\section{Analytical expressions of bifurcation curves for the APF case (Sec.~\ref{apf:sub}) }\label{app2}
We derive the expressions of Hopf bifurcation curves of the system with local APF given by \eqref{apf} using the same method as in Sec.~\ref{sec:lpf}. The derived expressions are as follows:
\begin{equation}
\alpha _{HB1,3}=\frac{-B_{HB3}\mp\sqrt{B_{HB3}^{2}-4A_{HB3}C_{HB3}}}{2A_{HB3}},
\end{equation}
where $A_{HB3}=\epsilon(1-Q)-2$, $B_{HB3}=(\epsilon Q+2)^2+\epsilon(2-\epsilon)$, and $C_{HB3}=-\epsilon^2(1+Q)^2-2(1+\omega^2)-\epsilon(1+Q)(3+\omega^2)$. $\mp$ sign is vertically aligned for HB1 and HB3.
\begin{equation}
\alpha _{HB2}=\frac{-B_{HB2}+\sqrt{B_{HB2}^{2}-4A_{HB2}C_{HB2}}}{2A_{HB2}},
\end{equation}
where
\begin{align*}
A_{HB2}&=2+\epsilon(1-Q)-\frac{8\omega^2}{L_{HB2}},\\
B_{HB2}&=\epsilon^2(Q^2-1)+4+\frac{4\omega^2(\epsilon-8+4Q\epsilon)}{L_{HB2}}\\
&+\frac{64\omega^4}{L_{HB2}^2}-4Q\epsilon-2(L_{HB2}-2\omega^2),\\
C_{HB2}&=\bigg[\omega^2+\frac{12\omega^4}{L_{HB2}^2}+\frac{2\omega^2[(\epsilon+2)(1+Q)-2]}{L_{HB2}}\\
&+2(1+Q)(\omega^2-\epsilon)\bigg](A_{HB2}-2\epsilon),\\
L_{HB2}&=\epsilon+\sqrt{\epsilon^2-4\omega^2}.
\end{align*}
And finally,
\begin{equation}
\alpha _{HB4}=\frac{-B_{HB4}+\sqrt{B_{HB4}^{2}-4A_{HB4}C_{HB4}}}{2A_{HB4}},
\end{equation}
where $A_{HB4}=2+\epsilon(1-Q)-\frac{8\omega^2}{L_{HB4}}$, $B_{HB4}=\epsilon^2(Q^2-1)+4+\frac{4\omega^2(\epsilon-8+4Q\epsilon)}{L_{HB4}}+\frac{64\omega^4}{L_{HB4}^2}-4Q\epsilon-\frac{2(L_{HB4}-2\omega^2)}{(1-Q)}$, $C_{HB4}=(A_{HB4}-2\epsilon)[\frac{\omega^2(Q+3)}{(1-Q)}-\frac{(1+Q)L_{HB4}}{(1-Q)}+\frac{12\omega^4}{L_{HB4}^2}+\frac{2\omega^2(\epsilon Q+\epsilon-2)}{L_{HB4}}]$, $L_{HB4}=\epsilon(1-Q)+\sqrt{\epsilon^2(1-Q)^2-4\omega^2}$.
\providecommand{\noopsort}[1]{}\providecommand{\singleletter}[1]{#1}%
\end{document}